# Constructing and evaluating machine-learned interatomic potentials for Li-based disordered rocksalts

Vijay Choyal, Nidhish Sagar, and Gopalakrishnan Sai Gautam*

Department of Materials Engineering, Indian Institute of Science, Bengaluru 560012, Karnataka, India

*E-mail: saigautamg@iisc.ac.in

## Abstract

Lithium-based disordered rocksalts (LDRs), which are an important class of positive electrode materials that can increase the energy density of current Li-ion batteries, represent a significantly complex chemical and configurational space for conventional density functional theory (DFT)-based high-throughput screening approaches. Notably, atom-centered machine-learned interatomic potentials (MLIPs) are a promising pathway to accurately model the potential energy surface of highly-disordered chemical spaces, such as LDRs, where the performance of such MLIPs have not been rigorously explored yet. Here, we represent a comprehensive evaluation of the accuracy, transferability, and ease of training of five atom-centered MLIPs, including the artificial neural network potentials developed by the atomic energy network (AENET), the Gaussian approximation potential (GAP), the spectral neighbor analysis potential (SNAP) and its quadratic extension (qSNAP), and the moment tensor potential (MTP), in modeling a 11-component LDR chemical space. Specifically, we generate a DFT-calculated dataset of 10842 configurations of disordered $LiTMO_2$ and $TMO_2$ compositions, where TM = Sc, Ti, V, Cr, Mn, Fe, Co, Ni, and/or Cu). Importantly, we find AENET to be the best potential, in terms of accuracy and transferability for energy predictions, while MTP is the best for atomic forces. While AENET is the fastest to train among the MLIPs considered at low number of epochs (300), the training time increases significantly as epochs increase (3300), with a corresponding reduction in training errors (~60%). Note that AENET and GAP tend to overfit in small datasets, with the extent of overfitting reducing with larger datasets. Finally, we observe AENET to provide reasonable predictions of average Li-intercalation voltages in layered, single-TM $LiTMO_2$ frameworks, compared to DFT (~10% error on average). Our study should pave the way both for discovering novel disordered rocksalt electrodes and for modelling other configurationally complex systems, such as high-entropy ceramics and alloys.

## Introduction

Lithium ion batteries (LIBs) are the workhorse energy storage technology that is powering modern portable electronics and electric vehicles, and remains a crucial ingredient in our transition to a carbon-free society.[1–3] Among the various components that constitute LIBs, the cathode or the positive electrode, which is usually made of lithium, combinations of few transition metals (TMs), and oxygen resulting in a $LiTMO_2$ composition,[4,5] is most critical in determining the energy density and cost of the overall electrochemical cell.[1,6,7] Thus, increasing the performance of cathode materials is an ongoing and active area of research. One pathway to increase the energy density of LIB cathodes is to utilize anionic redox (i.e., oxygen redox), which can be reversibly accessed in cathode materials with a Li-excess composition, i.e., $Li_{1+x}TM_{1-y}O_2$, where x and y are positive numbers less than 1.[8,9] Importantly, cathode frameworks that adopt a Li-excess composition along with a fluorinated disordered rocksalt structure (i.e., DRX cathodes) have displayed significant promise, with reversible and facile anionic redox, high voltages with minimal volume changes during charge/discharge, and macroscopic Li-ion transport via percolating networks.[10–21] Thus, designing and optimizing DRX cathodes can significantly improve the energy density and (potentially) cost of LIBs.[22]

The structure of a conventional $LiTMO_2$ cathode, such as $LiCoO_2$, is a layered framework, with Li-ions (green polyhedra) and TM-ions (blue polyhedra) occupying distinct layers along the *c*-axis, as shown in **Figure 1a**. Such layered frameworks consist of a face centered close-packed anionic lattice (red spheres indicate anions in **Figure 1**), with Li and TM cations occupying octahedral voids of adjacent close-packed {111} planes, which results in distinct Li and TM layers.[5] Thus, the layered $LiTMO_2$ cathode is a specific ordering of the rocksalt structure, created by the anionic (O) and cationic (Li+TM) sub-lattices. Note that the Li and TM ordering within a layered cathode can undergo disordering, i.e., large fractions of Li sites can be occupied by TM and vice-versa, resulting in the breaking of long-range order, as displayed in **Figure 1b**.[8] Note that disordering can occur in cathodes with multiple TMs also (see **Figure 1c**), and subsequent topotactic Li deintercalation can result in a disordered $TMO_2$ configuration as well (**Figure 1d**). However, disordered cathodes with a stoichiometric Li content (i.e., $LiTMO_2$) do not show appreciable electrochemical activity,[23–25] since there are not enough connected Li-conducting pathways available within the structure for macroscopic Li transport,[26] prompting the use of Li-excess compositions.[10] Nevertheless, $LiTMO_2$ remains the base or stoichiometric composition around which the Li, TM, and anionic content (e.g., fluorination) is changed for designing energy-dense DRX.

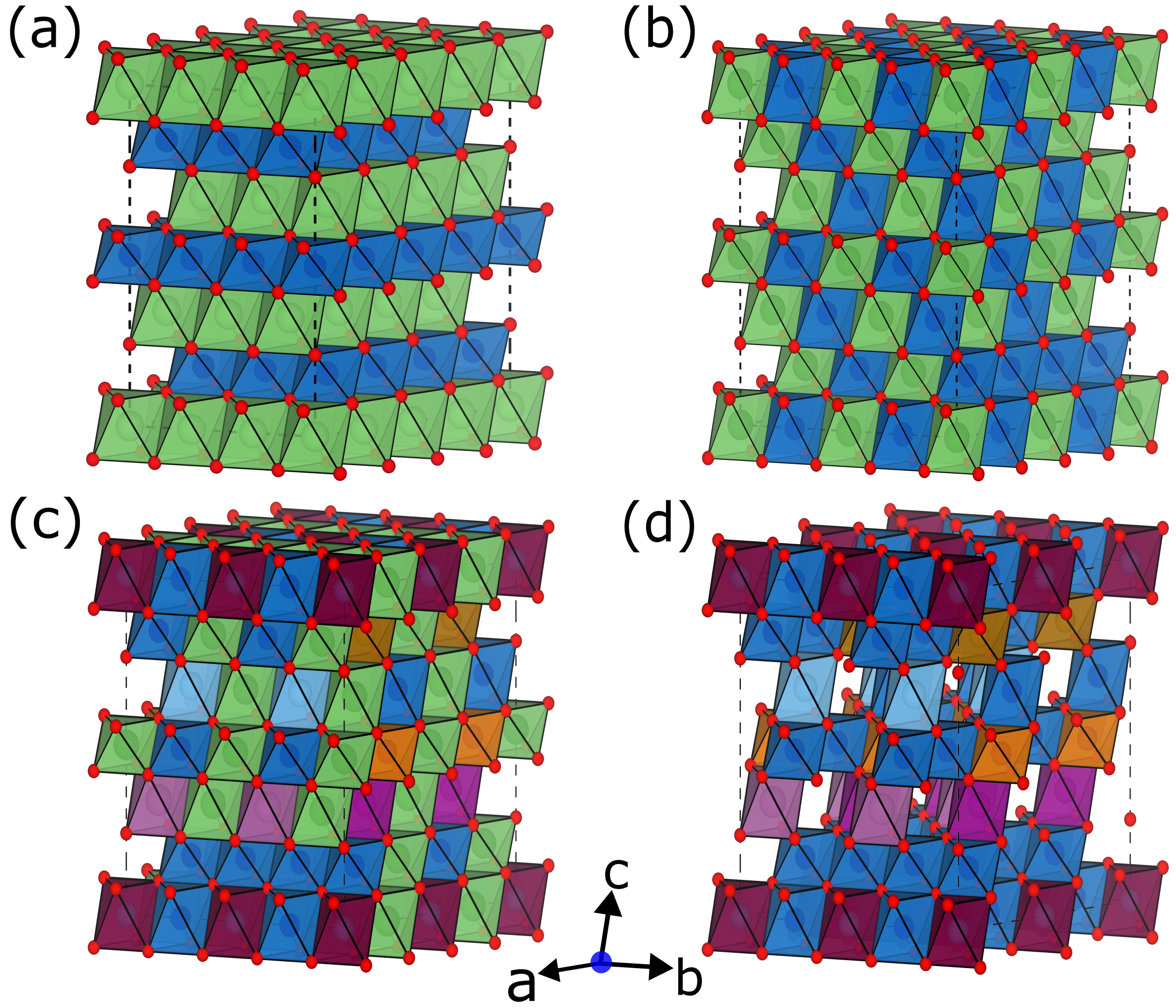

**Figure 1.** Structural schematic of (a) layered $LiTMO_2$, (b) disordered, single-TM $LiTMO_2$, (c) disordered multi-TM $LiTMO_2$, and (d) disordered, delithiated, multi-TM $TMO_2$. Green spheres/polyhedra indicate Li, red spheres indicate O, and TMs are indicated by other colors of spheres/polyhedra.

Off late, materials screening approaches using high-throughput density functional theory (DFT[27,28]) calculations with/without machine learning (ML), have resulted in a significant number of theory-predicted candidates in a variety of applications, including batteries.[10,29–42] Indeed, some of the theoretical predictions have also been validated by subsequent experiments.[43–49] However, modelling disordered rocksalt compositions, and subsequently performing a computational screening across various compositions is non-trivial, owing to the configurational complexity and length-scale of the system. Specifically, disordered rocksalts do not have significant long-range order, necessitating large supercells, which in turn results in multiple symmetrically distinct Li-TM arrangements to consider. The computational complexity becomes particularly severe in case of disordered structures containing multiple TMs, which is usually the case for several of the Li-excess cathodes that show good performance.[8] Thus, a conventional high-throughput screening approach is not practical for new DRX discovery, and this also applies for the general field of high entropy ceramics.[50,51]

An alternative method to model disordered multi-component systems such as disordered rocksalt cathodes, is to employ the so-called machine-learned interatomic potentials (MLIPs), which can provide 'quick' and 'accurate' estimates of energies and atomic forces within a given disordered configuration.[52–54] Specifically, MLIPs are usually trained on a (smaller) DFT-based training set and act as a mathematical approximation of the underlying potential energy surface (PES)[54] of the chemical system under consideration. Conventionally, MLIPs have been constructed on an atom-centered basis, i.e., the MLIP fingerprints the local environment around an atom-of-interest, and are designed to ensure invariance with rotation, reflection, translation, and permutations of the underlying atoms. Once constructed and validated, an MLIP can be used for larger-size and longer-time-scale simulations.

Examples of atom-centered MLIPs include the artificial neural network (ANN) potential[55,56] developed as part of the atomic energy network (AENET) package,[33,57,58] the gaussian approximation potential (GAP),[54,59,60] the moment tensor potential (MTP),[61–63] the spectral neighbor analysis potential (SNAP), and its quadratic version (qSNAP).[64,65] Training such potentials typically involves the generation of a sufficiently diverse dataset (usually with DFT), minimizing the root mean squared errors (RMSEs) and/or the mean absolute errors (MAEs) against target metrics (such as total energies and atomic forces), and optimization of relevant hyperparameters. These MLIPs are a significant improvement over classical forcefields, in terms of displaying near-DFT accuracy and being sufficiently quick in predicting energies, forces, and stresses.[33,66–69] While the performance of these potentials has been compared for single-component systems,[52] a rigorous benchmarking and testing of these potentials on a disordered system with a large number of components is missing so far.

Here, we perform a comprehensive analysis of the accuracy, transferability, and ease of training of five MLIPs, including MTP, SNAP, qSNAP, GAP, and AENET, on a highly diverse, 11-component dataset involving disordered $LiTMO_2$ compositions (TM = Sc, Ti, V, Cr, Mn, Fe, Co, Ni, and/or Cu). To train and test the potentials, we generate a DFT-calculated dataset of 10842 configurations involving different compositions of $LiTMO_2$, which was divided into training and testing data sub-sets of varied sizes. Subsequently, we quantify the accuracy via errors on energies and forces within training datasets, transferability via the similarity of errors across the training and corresponding testing sets, and ease of training through the computational training time required for each potential. Importantly, we find that AENET provides the best accuracy and transferability for total energies and is the easiest to train for a low number of epochs, while MTP is the best for atomic forces, exhibits robust transferability for total energies, and is difficult to train, highlighting that different MLIPs may be suited for different target metrics/applications. To further probe the accuracy of AENET on derived properties, we benchmark the potential-calculated average Li intercalation voltages versus DFT-estimates in ordered, layered, single-TM $LiTMO_2$ compositions (TM = Ti, V, Cr, Mn, Fe, Co, Ni, or Cu), with AENET showing a reasonable MAE~0.34 V (~10%), across all systems. We hope that our study and the potentials we have constructed provide new avenues for discovering novel DRX cathodes for LIBs, and a framework for screening through computationally complex, disordered, multi-component systems.

## Methods

### Workflow

An overview of the data generation, calculations employed, training of MLIPs, and quantifying predictive errors is shown in **Figure 2**. DFT total energies and atomic forces of six categories of training structures, i.e., structures involving single or multiple TMs, with or without Li, and ordered or disordered configurations, were calculated to create the training (and test) datasets (see **Dataset generation** subsection for additional details). A 90:10 random split of the dataset was used as training and test datasets, respectively, for the construction of all MLIPs. Note that the same random split(s) of structures was used for training/testing of all MLIPs. Subsequently, atomic descriptors (see **Table 1** and **Constructing MLIPs** subsection) were generated for the structures in the training set, which were used to train the MLIPs with the DFT total energies and/or atomic forces being the target properties. The hyperparameters for each MLIP framework were optimized during the training process to provide the minimum RMSE and MAE with respect to DFT-calculated energies and/or forces and are provided in **Table S1** of the supporting information (SI). Among the trained MLIPs, we chose the best MLIP (i.e., the AENET potential), which gave the lowest RMSE in energies and performed a comparison of the model-predicted Li-intercalation voltages versus DFT calculations in ordered, layered, single-TM $LiTMO_2$-$TMO_2$ compositions.

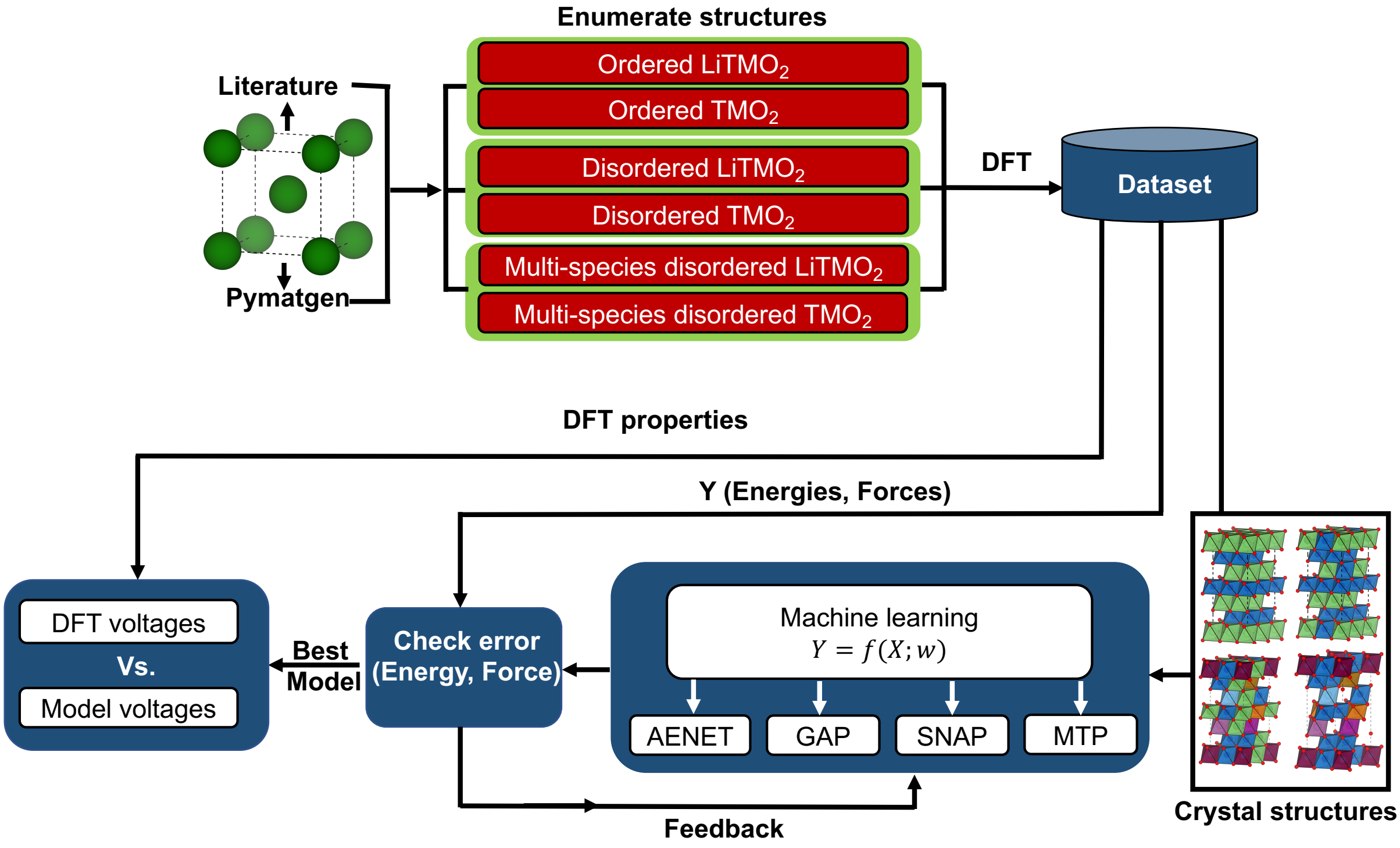

**Figure 2.** Workflow of structure enumeration, DFT calculations, MLIP training, and voltage predictions employed in this work.

## Dataset generation

The $LiTMO_2$ disordered rocksalt data set was generated by enumerating Li and TM arrangements within the cation sublattice of the rocksalt structure. Specifically, we used the structures enumerated by Artrith et al.[33] with various cation arrangements up to a total of 18 cation sites per cell using the enumeration approach of Hart et al.[70–72] Thus, Artrith et al.[33] generated a total of 10046 structures based on 9 TMs (i.e., TM = Sc, Ti, V, Cr, Mn, Fe, Co, Ni, and Cu). Subsequently, we generated specific combinations of TMs that were missing from the dataset of Artrith et al.[26] (e.g. $Li_9(Sc,Mn,V,Cr,Fe,Co,Cu,Ni)_9O_{18}$ and structures with all nine TMs) using pymatgen's advanced transformations module,[73] resulting in an additional ~400 structures. Also, we generated the structures without Li-ion to include Li-deficient conditions encountered at the top of charge. Specifically, we randomly selected a total of 500 structures from each of the $n$-TM combinations ($n$ = 1 to 9) and removed all Li atoms from such structures, bringing our final dataset to a total of 10842 structures.

## Constructing MLIPs

All MLIPs investigated in this work are atom centered, i.e., the MLIPs express the PES as a sum of atomic energies ($\epsilon_i$), where each atomic energy is a function of the local environment of each atom that obeys the underlying symmetry of the environment, such as rotational and translational invariance. The range of the local environment is determined by the cut-off radius ($R_c$ or $R_{cut}$) hyperparameter, with a corresponding switching function ($f_c$, a cosine function for example) that ensures that the contribution of neighboring atoms to the atomic energy decay smoothly to zero as the distance from the atom-of-interest approaches $R_c$. However, the MLIPs differ in the way the local environments are described mathematically and the functional basis/expression used to map the descriptors to the PES. The key concepts and model parameters used in all MLIPs considered are summarised in the following paragraphs and in **Table 1**, and can be found in the corresponding literature.[64,74,75]

**Table 1.** Summary of different MLIP frameworks considered in this work. SOAP and BFGS represent smooth overlap of atomic positions, and Broyden-Fletcher-Goldfarb-Shanno, respectively.

| MLIP | MTP | SNAP | GAP | AENET |
|---|---|---|---|---|
| **Idea** | Many-body interactions within a cut-off radius represented via moment tensors | Local atomic density (weighted delta functions) projected on a 4D hypersphere | Local atomic density modelled via a SOAP kernel | Feed forward neural network with local bonding environment as the input layer |
| **Descriptor** | Moment tensors consisting of radial distribution function and outer products of position vectors of neigboring atoms | Hyperspherical projection of atomic density expanded in terms of bispectrum components | Atomic density as weighted sum of Gaussians | Radial and angular distribution functions |
| **Training algorithm** | BFGS | Linear regression | Gaussian process regression | Limited memory BFGS |
| **Basis functions** | Chebychev polynomials (for radial basis) and contracted moment tensors | Hyperspherical harmonics | Equispaced Gaussians (for radial basis) and spherical harmonics (for angular basis) | Chebychev polynomials |

**MTP** represents the total energy of a configuration as a sum of $\epsilon_i$, which in turn is written as a linear expansion of basis functions, $B_\alpha^i$, with the linear coefficients ($\xi_\alpha$) determined by fitting to the training set. The atomic forces can be derived from MTP by taking spatial derivatives on the basis functions.

$$\epsilon_i = \sum_{\alpha} \xi_\alpha B_\alpha^i \tag{1}$$

The $B_\alpha^i$ are written using moment tensor descriptors ($M_{\mu,\nu}^i$), which consist of radial ($f_\mu$) and angular (outer products of position vectors of neighboring atoms, $j$, resulting in a tensor of rank $\nu$) components.[63] $r_{ij}$ is the distance between $i$ and neighbor $j$.

$$M_{\mu,\nu}^i = \sum_j f_\mu(r_{ij}) \underbrace{r_{ij} \otimes \cdots \otimes r_{ij}}_{\nu \text{ times}} \tag{2}$$

$f_\mu$ is expanded using a Chebychev polynomial basis of order $N_Q$ between a minimum ($R_{min}$) and $R_{cut}$ distance from $i$ to ensure smooth behavior of the MTP. The radial component has a set of radial parameters ($c_{\mu,i,j}$), which are fit to training data alongside $\xi_\alpha$. Once the $M_{\mu,\nu}^i$ are defined, each $B_\alpha^i$ is written as a "contracted" moment, where contraction of a moment can yield a scalar quantity from underlying vectors or tensors via dot or Frobenius products, respectively. Specifically, moments can be contracted to different "levels", depending on $\mu$ and $\nu$, with $B_\alpha$ (and eventually $\epsilon_i$) written up to a maximum level of contraction ($lev_{max}$). Thus, $N_Q, R_{min}, R_{cut},$ and $lev_{max}$ are crucial hyperparameters in the construction of MTP. Note that MTP is usually trained simultaneously on total energies, atomic forces, and stress tensors, with the relative weights given to these quantities during training being hyperparameters as well. The training algorithm that is typically employed by MTP is either Broyden-Fletcher-Goldfarb-Shanno (BFGS,[76] employed in this work) or linear regression.

**SNAP** describes the local atomic density $\rho_i(r)$ around an atom of interest, $i$, in terms of $\delta$ functions, as written below. $r_j$ is the position of a neighboring atom, $j$.

$$\rho_i(r) = \delta(r) + \sum_{r_j < R_{cut}} f_c(r_j) w_j \delta(r - r_j) \tag{3}$$

The density function is subsequently mapped on to three angular coordinates (instead of the usual two), and expanded via 4D hyperspherical harmonics.[65] Specifically, the distance coordinate $(r)$ is written in terms of a third angular coordinate, $\theta_0 = \frac{\theta_{max} r}{R_{cut}}$, in addition to the typical angular coordinates of a spherical system, $\theta$ and $\phi$, resulting in distinct points on a unit 3-sphere. Expansion using hyperspherical harmonics results in bispectrum components that are scalar triple products of the harmonic expansion coefficients, which in turn characterize the strength of density correlations. The hyperparameter $J_{max}$ determines the number of bispectrum components ($K$) to be used during the expansion. The $K$ bispectrum components, written arbitrarily as $B_1^i, B_2^i, \ldots, B_K^i$, in turn contribute to the energy of an individual atom ($\epsilon_i$) as a linear summation, weighted by $\beta_K$ coefficients.

$$\epsilon_i = \beta_0 + \sum_{k=1}^{K} \beta_k \left(B_k^i - B_{k0}^i\right) \tag{4}$$

$\beta_k B_{k0}^i$ is the contribution of each bispectrum component to the energy of an isolated $i$ atom, which results in $\beta_0$ being the energy of an isolated atom as calculated by SNAP. Atomic forces can be derived from **Equation 4** by obtaining spatial derivatives of $B_K^i$. The total energy of a given structure is written as a sum of atomic energies. Finally, the $\beta_k$ values are determined via linear regression with respect to training total energies and atomic forces. In the case of qSNAP, the formalism is identical to SNAP except for additionally including the contributions from all pairwise products of bispectrum components in **Equation 4**. Hence, the number of hyperparameters are identical for SNAP and qSNAP.

**GAP** uses a kernel based on the smooth overlap of atomic positions (SOAP[59,77]) to describe the atomic density in the neighborhood of the reference atom in combination with Gaussian process regression to evaluate atomic energies and forces. The $\rho_i$ at a distance $R$ from a central atom $i$ is described using a squared exponential function in SOAP, where $R_{ij}$ is the distance between a neighbor $j$ and $i$ within the $R_c$, and $\sigma_{atom}$ can be considered to represent the size of atoms, which controls the smoothness of the SOAP kernel.

$$\rho_i(R) = \sum_j f_c(R_c) \exp\left(-\frac{\left|R - R_{ij}\right|^2}{2\sigma_{atom}^2}\right) \tag{5}$$

$\epsilon_i$ can be represented as a weighted ($w(r)$) integral of the atomic density functions, where the weights follow a Gaussian distribution. The total energy of a structure, is in turn, the summation of the atomic energies. The atomic forces can be obtained via partial derivatives of the total energy.

$$\epsilon_i = \int w(r)\rho_i(r)dr \quad (6)$$

In practice, the SOAP kernel is written as a numerical expansion (**Equation 7**), typically involving equispaced Gaussians as the orthonormal radial basis functions ($g_n(r)$), and spherical harmonics as the angular basis functions ($Y_{lm}(\hat{r})$).[60] $c^i_{nlm}$ are the expansion coefficients of the atom $i$, which are determined using Gaussian process regression against a training dataset of energies and/or forces by evaluating their covariance matrix with other atoms $j$ in the structure. The hyperparameters $n_{max}, l_{max}, \zeta$ determine the maximum powers of the radial and angular components in the expansion basis, and the power factor to accentuate the sensitivity of the SOAP kernel, respectively.

$$\rho_i(r) = \sum_{nlm} c^i_{nlm}\ g_n(r)Y_{lm}(\hat{r}) \quad (7)$$

**AENET** employs a feed forward neural network architecture, where multiple artificial neurons are arranged in layers, with each neuron connected only to neurons in adjacent layers via activation functions. Similar to MTP, SNAP, and GAP, AENET also describes the energy of a given configuration as the sum of atomic energies, with atomic forces calculated using spatial derivatives of the total energy. Importantly, AENET fingerprints the local environment of each atom $i$ using radial and angular distribution functions ($RDF_i$ and $ADF_i$), which are expanded using Chebychev polynomials. Notably, the $RDF_i$ and $ADF_i$ do not increase the computational complexity of the ANN with number of species in the structure as was the case in the classical Behler-Parinello ANNs.[55,56]

$$RDF_i(r) = \sum_{R_{ij}<R_{cut}} \delta(r - R_{ij})f_c(R_{ij})w_{t_j}$$

$$ADF_i(\theta) = \sum_{R_{ij},R_{ik}<R_{cut}} \delta(\theta - \theta_{ijk})f_c(R_{ij})f_c(R_{ik})w_{t_j}w_{t_k} \quad (8)$$

The weights, $w_{t_j}$ and $w_{t_k}$ are assigned values such that there are a unique set of Chebychev expansion coefficients describing the structural features and another set of coefficients that describe atom types (i.e., four sets of expansion coefficients, two each of radial and angular expansion coefficients). Subsequently, the $RDF_i$ and $ADF_i$ constitute the input node to the ANN, with the output node being the atomic energy of species $i$. The weights and biases within the ANN are obtained based on the training set of energies, via the limited memory-BFGS algorithm.[60,78] The important hyperparameters for training the AENET potential include the order of the Chebychev basis expansion used ($N$), number of hidden layers and nodes per layer in the neural network, the cut-off radii for $RDF_i$ and $ADF_i$, and the number of epochs (i.e., number of training iterations) employed during training.

## DFT Calculations

We performed Hubbard $U$ corrected[79] DFT calculations using the Perdew-Burke-Ernzerhof (PBE) functionalization of the generalized gradient approximation (GGA),[80] with projector-augmented wave (PAW) potentials,[81,82] as implemented in the Vienna ab initio simulation package (VASP, version 6.1.2).[83,84] The calculations were performed, without preserving any symmetry, for all the 10842 structures till total energies converged to within 0.01 meV. Note that we performed a single self-consistent field calculation for the converged structures obtained from Artrith et al.,[33] while we performed a full structure relaxation (i.e., relax ionic positions, cell volume, and cell shape) for all the structures we enumerated and for Li-intercalation voltage calculations. For all structures that were fully relaxed, we converged both the total energies and atomic forces to within 0.01 meV and 30 meV/Å, respectively.

Γ-centred $k$-point meshes with a density of 1000 divided by the number of atoms were used, in accordance with the work of Artrith et al.[33] The plane wave kinetic energy cutoff was set to 520 eV and VASP input parameters were generated using pymatgen,[73] where the parameters were compatible with the Materials Project.[85] The reference atomic energies in the AENET package were calculated by placing isolated Li, TM, and O atom, at the origin of a 18×19×20 $Å^3$ cell (i.e., each reference energy was calculated for an isolated atom). **Table S2** summarizes Hubbard $U$ values and the corresponding calculated atomic energies. We used the materials machine learning (MAML[52]) python package as an interface to construct the GAP, SNAP, qSNAP, and MTP potentials. For voltage calculations, we considered layered, single-TM $LiTMO_2$ structures obtained from the inorganic crystal structure database (ICSD[86]) and the corresponding delithiated versions, except $LiTiO_2$, where we obtained the structure via ionic substitution of layered-$LiVO_2$. While DFT voltage values were obtained from a full structure relaxation of $LiTMO_2$ and $TMO_2$ structures, we obtained AENET voltages using the corresponding initial structures used for DFT calculations.

## Results

### Training and test errors

The RMSE in energies (in units of meV/atom) and forces (in eV/Å) errors for all the four MLIPs are summarized in the top and bottom panels of **Figure 3**, respectively, as a function of increasing total dataset size, with the parity plots between MLIP- and DFT-calculated values compiled in **Figures S2-S11**. Specifically, we display the progression of errors for total dataset sizes from 2000 till 9000 in steps of 1000, and for the full dataset of 10842 configurations, where each dataset is split into a 90:10 training:test set. Note that each smaller dataset is a randomly chosen subset from the 10842 structures, thus including the diversity and complexity of the local atomic environments to an extent. Each row of **Figure 3** represents an MLIP, while the top and bottom triangles within each box represent the errors associated with the training and test sets, respectively, for each total dataset size. The errors displayed in **Figure 3** correspond to the best set of hyperparameters that we identified for each MLIP (see **Table S1**). We do not include qSNAP error data in **Figure 3** since we observed qSNAP to provide similar metrics as SNAP when trained on small datasets (i.e., 2000 and 3000, see **Figure S1**), indicating marginal improvement over SNAP. Moreover, we encountered numerical and convergence difficulties with qSNAP while attempting to train it on larger dataset sizes. Hence, we do not include qSNAP in comparison to the other MLIPs for the rest of this study.

In general, the MLIPs considered show higher training errors in energy with an increase in training dataset size (lower triangles in **Figure 3**), although the increase is not always monotonous and the extent of increase is different for different potentials. For example, the training set energy errors increased from 21.89 to 23.04 meV/atom in MTP, 78.77 to 82.79 meV/atom in SNAP, 1.49 to 17.11 meV/atom in GAP, and 0.85 to 7.51 meV/atom in AENET. While AENET and GAP display largely monotonic increase in error with dataset size, indicating that these frameworks become better fitted with more data, MTP and SNAP are non-monotonic and show relatively low increase (~5%) in error as the total dataset size is increased from 2000 to 10842. Thus, the choice of the dataset itself may play a role in the accuracy of MTP and SNAP fits and more data may not necessarily improve the fit. Importantly, AENET exhibits the lowest error with our full training dataset, indicating high accuracy during training. Indeed, AENET's RMSE of ~7.5 meV/atom is not far away from the typical error expected within DFT calculations (~1 meV/atom). In contrast, SNAP shows the highest error across all dataset sizes, which may be due to the high number of components in our dataset.

Different from training errors on energies, test errors (upper triangles in **Figure 3**) monotonically decrease with increasing dataset size for all MLIPs, except MTP, which indicates an improvement in the transferability of all models with increasing training dataset size. Additionally, the MLIPs do display higher test errors compared to training errors, which is expected, except for MTP and SNAP, which show lower test than training errors for larger datasets. Notably, GAP and AENET display significantly high test errors compared to training, especially in small datasets (<6000 datapoints), which is indicative of a high degree of overfitting. For example, GAP's test error is 122 meV/atom compared to a training error of 1.49 meV/atom for a total dataset size of 2000, while AENET's test and training errors for the same dataset size are 49.9 and 0.85 meV/atom, respectively. Thus, both AENET and GAP are susceptible to overfit to the available data, especially when the dataset size is of the order of few thousand points or below. However, both AENET and GAP models become more transferable with increase in the overall dataset size, as exhibited by similar training and test errors as the dataset size approaches 10842. The improvement in the accuracy and transferability of AENET is consistent with prior observations in literature using ANNPs as well.[87,88] Interestingly, both MTP and SNAP show similar training and test errors across all dataset sizes (<11 meV/atom deviation) indicating good transferability of MTP and SNAP, though the accuracy of these MLIPs may not be as good as AENET or GAP. Thus, for the full dataset (10842 points), AENET gives both best accuracy and transferability for total energy evaluations, while MTP provides best transferability for smaller datasets (< 7000 points).

Compared to training and predicting total energies, all MLIPs do display significantly larger errors in training or predicting atomic forces, as compared to DFT (lower panel of **Figure 2**), which is expected given that force data is usually noisier compared to energy data. Indeed, the lowest force training error that we encountered is 0.21 eV/Å by MTP for the 3000 dataset, which is at least an order of magnitude higher than the errors typically encountered in DFT calculations (~0.03-0.05 eV/Å). In terms of training the MLIPs, AENET does not yet have a provision to weight forces during training, while we did not explicitly include forces during the training of GAP to lower computational time and limit memory usage. Consequently, AENET and GAP show significantly high training errors on atomic forces, compared to MTP and SNAP. For example, AENET's RMSE on forces in the 10842 training set is 1.10 eV/Å, while GAP displays an RMSE of 0.87 eV/Å, significantly higher than MTP or SNAP (0.22-0.24 eV/Å). While the force training errors on GAP can be certainly brought lower by a careful optimization of the "default_sigma" hyperparameter that controls the relative weights between forces and energies apart from the regularity of the fit,[75] AENET's training algorithm requires an extension if force errors need to be brought down systematically for any dataset.

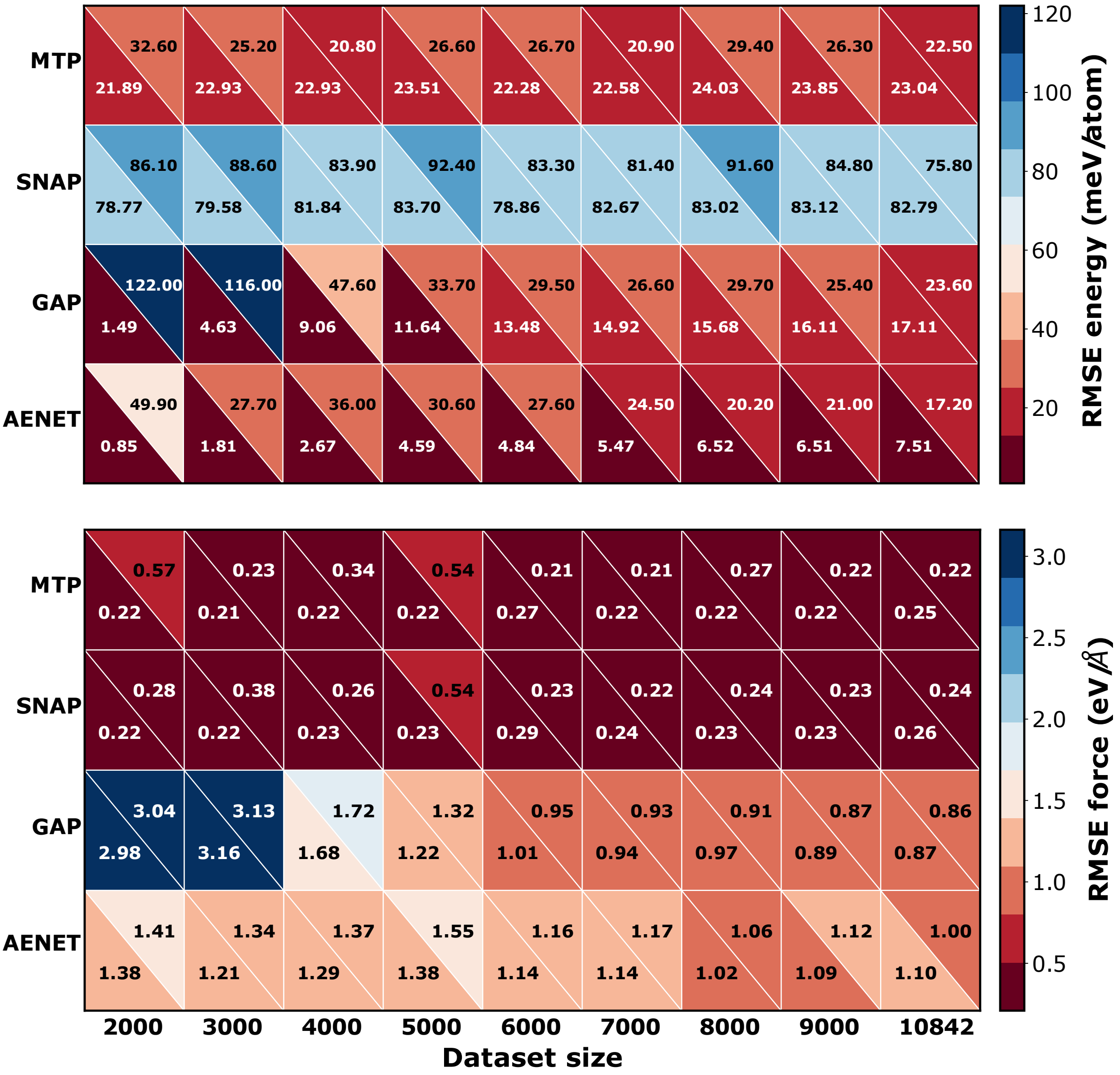

**Figure 3.** RMSE in energies (top panel) and forces (bottom panel) for the MLIPs considered in this work, as a function of increasing total dataset size. Lower left and upper right triangles within each square represent training and test errors, respectively.

All MLIPs, except AENET, display good transferability in terms of force predictions, given the similarity of training and test force errors, particularly at large dataset sizes. For example, MTP, SNAP, and GAP display <0.03 eV/Å deviation within training and test errors for the 10842 dataset. Thus, given both energy and force errors during training and their transferability to the test dataset, we find MTP to be the best performer, with marginal improvement in accuracy with increasing dataset size, and most suited for any dynamics simulations among the MLIPs considered. However, in terms of energy training and transferability, we find AENET to be the best performer, with a systematic improvement in accuracy and transferability with increasing data, and the most suited for any quick evaluation of total energies for a given structure, i.e., for static calculations. Note that the overall utility of a given MLIP is also determined by the computational time taken for training, which is discussed in the following section.

## Ease of training

The computational time taken for training the MLIPs as a function of increasing total dataset size (from 2000 to 10842), for the set of optimal hyperparameters (**Table S1**), is displayed in **Figure 4**. We used a single core of a Intel® Xeon® Gold 6271 CPU, with a maximum random access memory of 128 Gigabytes, without any hyperthreading for training all MLIPs. For AENET, the computational training time plotted in **Figure 4** is for 300 epochs, while we have performed an extended analysis on the dependence of the training time on number of epochs later this section (see **Figure 5**). As expected, all MLIPs require larger training times with increasing dataset size. Importantly, we find AENET (blue diamonds in **Figure 4**) to be the swiftest to train for all dataset sizes at 300 epochs, except 2000 dataset size, where SNAP (purple squares) is the fastest. Indeed, AENET is more than twice as fast (~210 min) compared to MTP (~460 min). Also, we find MTP (red triangles) to be the slowest to train for all dataset sizes, followed by GAP (green circles).

Moreover, MTP and GAP both become progressively harder to train with increasing dataset sizes. For example, MTP and GAP require an additional ~37 and ~28 min per additional 1000 total datapoints included, respectively, while AENET and SNAP are significantly easier to train, requiring additional times of ~15 and ~20 min per additional 1000 datapoints. Thus, the marginal improvement in accuracy on energies and forces with increasing datapoints exhibited by MTP is counterweighed, to an extent, by the significantly higher computational times required to train the potential. Note that we did not explicitly include forces during the training of GAP and we expect the computational training time of GAP to significantly increase, if forces were to be included as well. Therefore, we find that AENET is the easiest to train at 300 epochs, compared to the other MLIPs considered in this work.

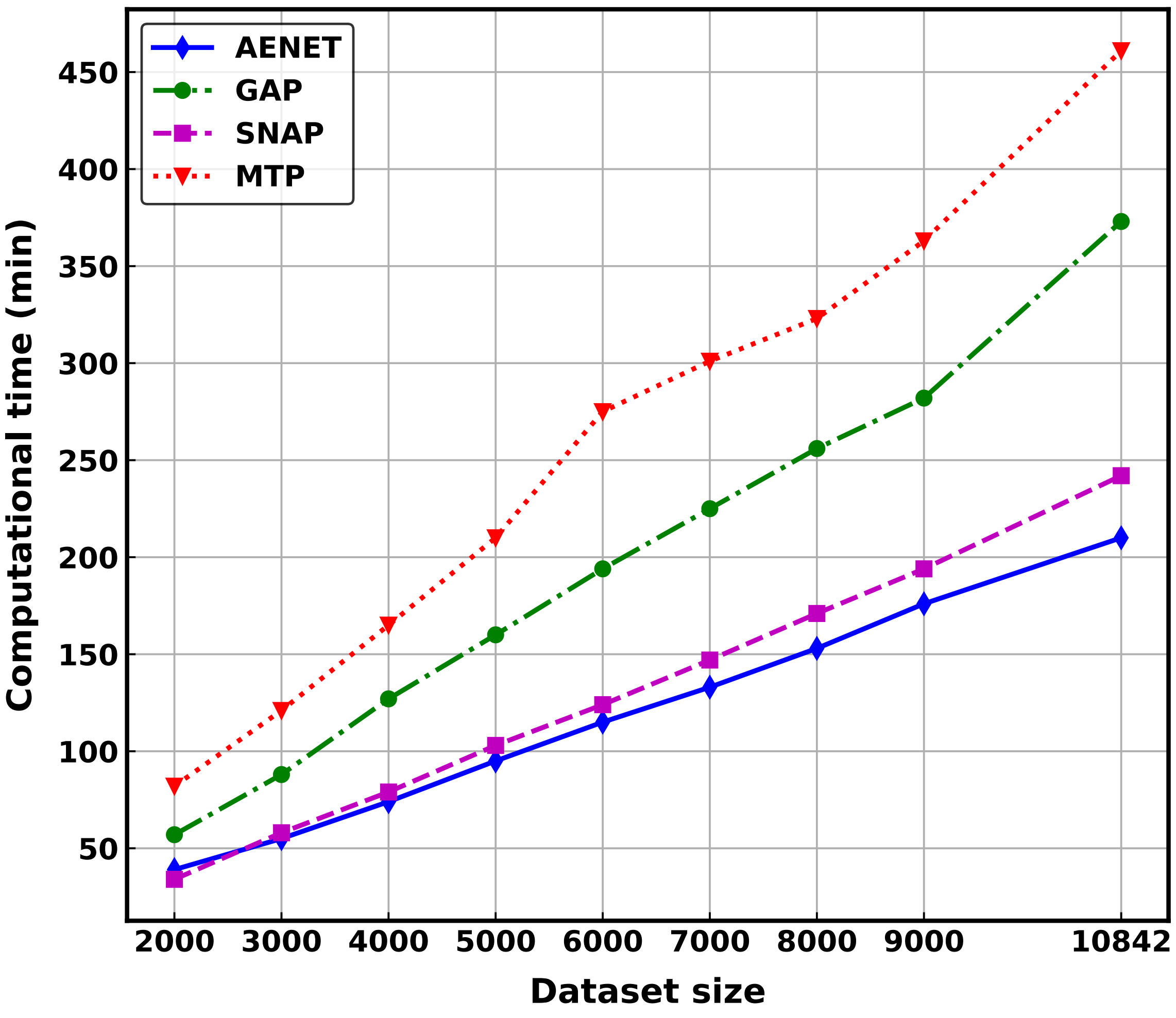

**Figure 4.** Computational time (in minutes) for training versus total dataset size for all MLIPs considered in this work.

Note that one of the key hyperparameters that influences (training) errors, and computational time in AENET is the number of epochs used during training. The AENET results displayed in **Figures 3** and **4** correspond to 3300 epochs of training. To further probe the role of this hyperparameter, we examine the variation of the computational training time as a function of increasing total dataset size and number of epochs (**Figure 5a**) and the change in training RMSE and computational training time for the full 10842 dataset with increasing epochs (**Figure 5b**).

For all dataset sizes, we observe an increase in computational time with increasing epochs (**Figure 5a**). More importantly, the rate of increase in computational time is more rapid with increasing epochs for larger datasets, indicating that the training time can quickly escalate as the complexity of the training set increases. For example, the training time required for 3300 epochs over the full 10842 dataset is ~38 hours (or ~2280 min, **Figure 5a**), which is ~5× the training time required by MTP for the full dataset (**Figure 4**). However, we do note that the energy RMSEs for the training set decreases steadily with increasing epochs (**Figure 5b**), from ~18.7 meV/atom at 300 epochs to ~7.5 meV/atom at 3300 epochs (i.e., a ~60% drop in errors). Nevertheless, the rapid increase in computational time for the full 10842 dataset may not be worth the marginal gain in accuracy, especially from ~27 hours, ~8.5 meV/atom at 2300 epochs to ~38 hours, ~7.5 meV/atom at 3300 epochs, or a ~41% gain in training time for an increase in accuracy of ~12% (**Figure 5b**). Therefore, AENET can be a quick MLIP to train compared to MTP, GAP, or SNAP, only if a low number of epochs gives reasonable accuracy and transferability for total energies.

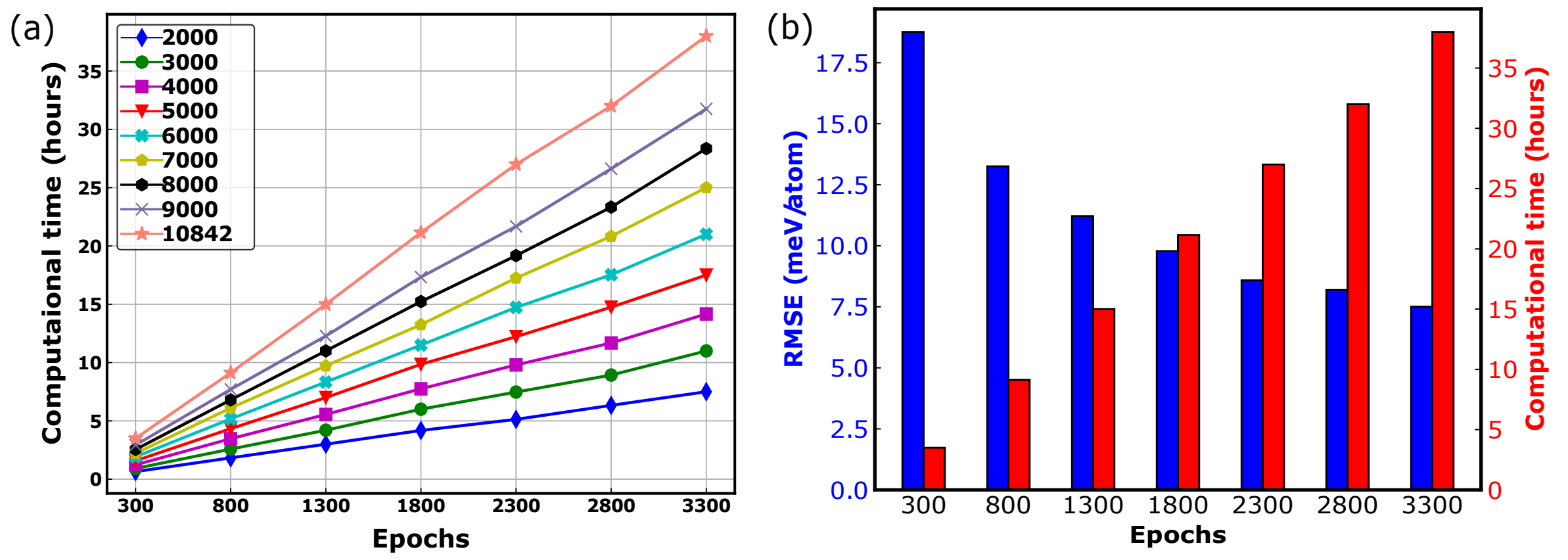

**Figure 5.** (a) Computational training time (in hours) versus number of epochs for different total dataset sizes using AENET. (b) Training energy RMSE (left y-axis and blue bars) using AENET for the total 10842 dataset with increasing number of epochs, with the corresponding computational training time (in hours) also indicated (right y-axis and red bars).

## Voltage predictions

The accuracy in predicting material properties is critical for evaluating the performance and determining the utility of MLIPs. Hence, we choose the prediction of average intercalation voltages, in ordered, layered, single-TM $LiTMO_2$ compounds (TM = Ti, V, Cr, Mn, Fe, Co, Ni, or Cu) as a test of accuracy, versus DFT calculations. Specifically, we examine the accuracy of the AENET potential in determining voltages, since it provided the best accuracy and transferability in prediction energies within disordered systems (see **Figure 3**). For voltage calculations, we utilize the AENET potential trained at 2300 epochs (since training errors declined by only ~1.0 meV/atom from 2300 to 3300 epochs, see **Figure 5b**). Note that the average Li intercalation voltage, versus Li metal, can be approximately determined using the following relation,

$$V = -\left(\frac{E_{\mathrm{LiTMO_2}} - E_{\mathrm{TMO_2}} - E_{\mathrm{Li}}}{F}\right) \quad (9)$$

where $E_{\mathrm{LiTMO_2}}$, $E_{\mathrm{TMO_2}}$, and $E_{\mathrm{Li}}$ are the total energies per formula unit of the lithiated structure, delithiated compound, and pure Li, respectively, from DFT or AENET, with the $p - V$ and entropic contributions ignored. $F$ is the Faraday's constant. Note that we used the DFT calculated total energy for Li in case of AENET predictions also, since the energy scales of our fitted AENET potential were quite different from the energy scale of a pure Li in its body centred cubic ground state configuration. We performed either structure relaxation using DFT or a static calculation using AENET, from the ICSD starting structures, to obtain the energies listed in **Equation 9**. All DFT and AENET calculated voltage values are presented in **Table S3** of the SI.

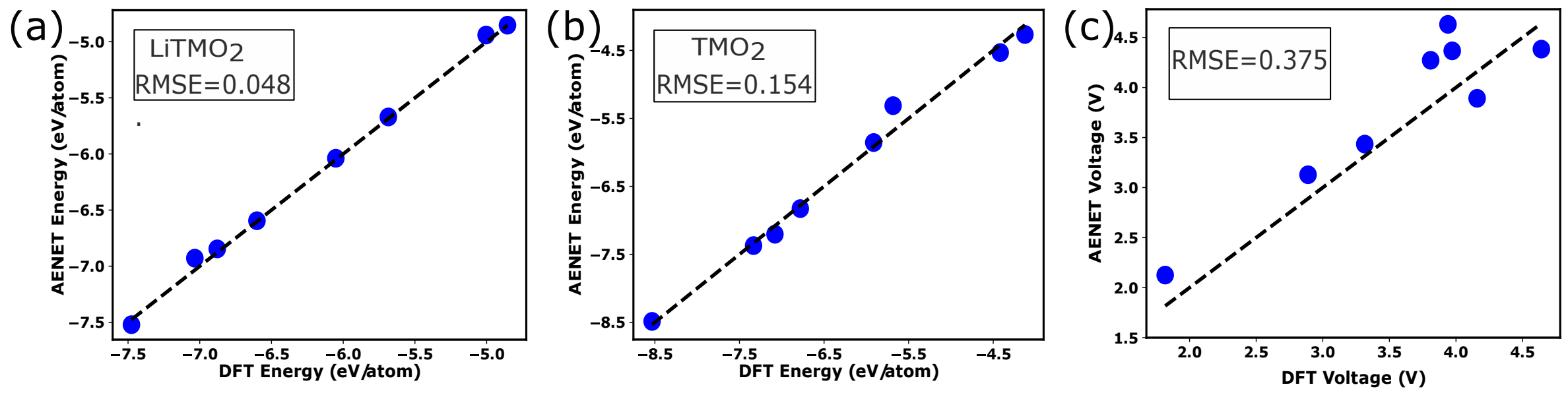

**Figure 6**. Comparison of AENET and DFT calculated energies of $LiTMO_2$ (panel a) and $TMO_2$ (panel b) compounds, where TM = Ti, V, Cr, Mn, Fe, Co, Ni, or Cu. Panel c represents a similar AENET-DFT comparison in terms of average Li intercalation voltage, versus Li metal.

**Figure 6** displays AENET calculated values versus DFT estimates, in terms of total energies (in eV/atom) of the fully lithiated compositions (panel a), fully delithiated compositions (panel b), the average voltages (panel c, in V vs. Li metal). The black lines in each panel of **Figure 6** indicate parity lines, with the degree of agreement between AENET and DFT values quantified in terms of RMSEs in each panel. Importantly, AENET estimates the total energies of both the lithiated and delithiated compositions quite accurately, with RMSEs of 0.048 eV/atom and 0.154 eV/atom (with corresponding MAEs of 0.035 eV/atom and 0.117 eV/atom), respectively, where the errors for lithiated compositions are similar to the test errors observed in **Figure 3**. These RMSEs represent percentage errors of 0.55% for $LiTMO_2$ and 2.12% for $TMO_2$ compositions. The higher errors within the $TMO_2$ compositions are expected since our dataset has significantly more $LiTMO_2$ configurations than $TMO_2$.

In terms of voltage calculations, AENET does suffer from a compounding of error (instead of a cancellation of error) versus DFT, with the RMSE at ~0.375 V (and MAE of

~0.342 V) compared to DFT for all systems. The voltage errors represent a percentage error of 10.05% against DFT, which is higher than the percentage errors on the energy estimates. Notably, AENET's largest voltage errors are in the $LiFeO_2$ and $LiTiO_2$ systems (~17%). Nevertheless, we expect our AENET potential to be useful for obtaining 'quick' voltage predictions for several disordered lithium transition metal oxide compositions, with reasonable accuracy, provided that the potential is not utilized for a full structure relaxation.

## Discussion

We have performed a comprehensive evaluation on the accuracy, transferability, and ease of training of four different MLIPs, namely AENET, GAP, SNAP, and MTP, on a 11-component lithium-containing disordered rocksalt chemical space (**Figure 1**), which is quite relevant for designing energy-dense cathodes in Li-ion batteries. Specifically, we have quantified the training and test errors (**Figure 3**) and the computational training time (**Figure 4**) of the above mentioned MLIPs on a DFT-calculated dataset of 10842 configurations, based on a computational workflow (**Figure 2**). While we found MTP to be accurate and transferable in terms of force predictions, we found AENET to be significantly more accurate and transferable for energy predictions. In terms of ease of training, we found AENET to be fastest to train at 300 epochs, with training time increasing ~10× at 3300 epochs and training errors decreasing by ~60% (**Figure 5**). While MTP is ~2× slower than AENET at 300 epochs, it is ~5× faster than AENET at 3300 epochs (**Figures 4** and **5**). We also observed AENET's voltage predictions at 2300 epochs (**Figure 6**) to be reasonable compared to DFT for layered $LiTMO_2$. Thus, we hope that our work propels the use of MLIPs to model highly disordered ceramic and/or metallic systems.

For the training of any MLIP on any chemical system, the following factors are important: (i) accuracy in terms of low errors on energy and force predictions (and any other relevant properties) on the training dataset, (ii) transferability, as quantified by the similarity of errors between the training and test datasets, and (iii) quality of the dataset, in terms of sampling a sufficient number of relevant local chemical environments, and (iv) computational cost, be it for generating the dataset or for training the MLIP itself. While these factors have a degree of dependence on one another, we found that an increase in dataset size generally lead to higher accuracy and better transferability, especially for force predictions, although the rate of reduction in errors begin diminishing at different rates for different MLIPs. For example, both MTP and SNAP do not show significant changes in their train/test force errors beyond a 7000 dataset size, while AENET and GAP do not display a similar behavior. Thus, increasing the dataset size is not always a solution for improving certain property predictions, while diversity of the local environments sampled within the dataset and the choice of a given MLIP itself may play a bigger role in improving predictions.

One of the overarching concerns while constructing MLIPs is the tendency for the frameworks to be overfitted, which is generally indicated by a significant difference in errors in the test versus training datasets. Importantly, we find that both AENET and GAP tend to overfit at small dataset sizes (<3000 data points, see **Figure 3**), with the tendency to overfit decreasing significantly at larger datasets. Note that the definition of a small dataset size that results in overfitting of AENET or GAP will be dependent on the underlying chemical system, especially on the number of components present. Indeed, previous studies on fewer component systems have shown remarkably good fits with AENET and GAP.[57,89–91] Nevertheless, we advocate a careful evaluation of the train and test errors with increasing dataset size to be done while using AENET or GAP to ensure appropriate fits. Surprisingly, we found MTP and SNAP to be quite resilient to overfitting and show remarkably similar training and test errors across all dataset sizes. However, the accuracy on energies for both MTP and SNAP may have been limited by the high number of components (11) considered in this study.

In terms of ease of training, we found AENET and MTP to be fastest and slowest potentials to train, across all dataset sizes (except 2000 datapoints, **Figure 4**). However, AENET is fastest only at a low number of epochs (300), which results in higher errors. Indeed, increasing the number of epochs does significantly slow AENET down (**Figure 5**), with marginal reductions in training/test errors. Thus, if AENET were to be used for other systems, the number of epochs is an important hyperparameter to be optimized to get the best tradeoff of computational time and accuracy.

For performing predictions that can be practically utilized for the design of battery systems, we examined the accuracy of our AENET potential for average voltage predictions in layered, single-TM $LiTMO_2$, as compared to DFT. While we found high accuracy in energy predictions of $LiTMO_2$ and $TMO_2$ ordered configurations using AENET, the accuracy does drop for voltage predictions due to compounding of error, especially in Fe and Ti containing systems. To further probe the quality of AENET's voltage predictions on disordered lithium containing rocksalt systems, more DFT calculations are required on chemical combinations and/or configurations that we have not sampled so far. While this is an immediate follow-up exercise that we intend to take, we do expect AENET to provide quick, qualitative voltage predictions reasonably well for disordered rocksalt configurations as well. Thus, high-throughput screening approaches to unearth novel, high-voltage disordered rocksalt cathode compositions may be feasible with AENET potentials.

Apart from dataset generation that was constrained by computational costs, another limitation of our study is that we have explored MLIPs that are atom-centered. For example, AENET and MTP are parameterizations of radial and angular distributions of atoms around a central atom of interest, while GAP and SNAP employ descriptors to quantify the local density around an atom. There are several graph-network-based (neural net) potentials that have been developed recently, such as SchNet,[92] NequIP,[93] MEGNet,[94] and CHGNet[95] with the work by Reiser et al.[96] providing a well-compiled summary of available graph-based MLIPs. Most graph network potentials can be combined with deep learning to provide high accuracy for multi-component systems, as illustrated by MEGNet[97] and CHGNet.[95] Long range interactions, such as electrostatics and dispersions, have also been included with such graph-based potentials.[98] However, incorporating deep learning often comes with the requirement of a "large" dataset, such as the Materials Project,[85] and large computational training time as well. Moreover, SchNet and NequIP have not been used for high-component (>5) chemical spaces, as far as we know, thus their performance remains unexplored.

On the other hand, deep learning can be combined with atom centered basis to generate highly accurate MLIPs for low-component systems, as characterized by the DeePMD[99] framework. Also, symbolic regression can be used to construct highly transferable potentials with minimal parameters.[100] Nevertheless, we believe that the potentials explored in this study are highly relevant for describing low-to-medium component systems where data is scarce, and are well suited for performing a large number of molecular dynamics simulations given their ease of training and computational speed during predictions. Going forward, developing interatomic potential frameworks that can deal with data scarcity in the broad field of materials science and are resilient against overfitting[101] will be of high utility in studying disordered and/or large scale systems over long times.

## Conclusion

In this work, we have quantified the accuracy, transferability, and ease of training of five atom-centered MLIPs, namely MTP, SNAP, qSNAP, GAP, and AENET, in their ability to model the PES of disordered, 11-component, $LiTMO_2$ compositions (TM = Sc, Ti, V, Cr, Mn, Fe, Co, Ni, and/or Cu), which form the base of designing energy-dense DRX cathodes for advanced LIBs. To train these potentials, we generated a DFT-calculated dataset of 10842 configurations and created a 90:10 random split of the set for training and testing. Using magnitude of training errors and similarity between training and test errors as the metrics of accuracy and transferability, respectively, we found AENET to be the best potential for predicting total energies, while MTP was the best performer for atomic forces. While both AENET and GAP tended to overfit in small datasets, the extent of overfitting and the transferability of these potentials improve considerably with increase in dataset size. In terms of ease of training, we observed AENET (MTP) to exhibit the smallest (largest) computational training time, where the number of epochs is low (~300) for AENET. However, note that AENET's training time does increase significantly with increasing the number of epochs during training, which may hamper swift training of AENET potentials in specific systems. Finally, we found AENET to also provide reasonable predictions of derived quantities, such as average Li-intercalation voltages, in layered, ordered, single-TM $LiTMO_2$ compositions, versus DFT estimates. We hope that our work inspires more studies on using atom-centered MLIPs to model configurationally complex systems, resulting in the discovery of new DRX cathodes and other high-entropy ceramic/metallic systems.

### Conflicts of interest

There are no conflicts to declare.

### Author contributions

V.C. and N.S. performed all DFT calculations and construction of MLIPs, under the supervision of G.S.G. V.C. wrote the initial draft of the paper. G.S.G edited and refined the draft. All authors have approved the final version of the manuscript.

**Acknowledgements**

G.S.G. acknowledges financial support from the Indian Institute of Science (IISc) Seed Grant, SG/MHRD/20/0020 and SR/MHRD/20/0013 and the Science and Engineering Research Board (SERB) of the Department of Science and Technology, Government of India, under sanction numbers SRG/2021/000201 and IPA/2021/000007. V.C. thanks the Institute of Eminence Post-doctoral fellowship awarded by the Indian Institute of Science for financial assistance. A portion of the density functional theory calculations showcased in this work were performed with the computational resources provided by the Supercomputer Education and Research Center, Indian Institute of Science. We acknowledge National Supercomputing Mission (NSM) for providing computing resources of 'PARAM Siddhi-AI', under National PARAM Supercomputing Facility (NPSF), C-DAC, Pune and supported by the Ministry of Electronics and Information Technology (MeitY) and Department of Science and Technology (DST), Government of India.

**Data and Code Availability**

All computed data, relevant scripts, and the best MLIPs constructed in this work are available freely for all via our GitHub repository: https://github.com/sai-mat-group/mlip-disordered-rocksalts